\documentstyle[12pt]{article}

\advance \topmargin by -\headheight
\advance \topmargin by -\headsep

\evensidemargin \oddsidemargin
\begin{document}

\newpage
\pagestyle{empty}

\centerline{\large\bf 
Rayleigh-Ritz Variational Approximation}
\centerline{\large\bf 
and Symmetry Nonrestoration}
\vskip 2 cm
\begin{center}
{\bf Giovanni AMELINO-CAMELIA}\\
\end{center}
\begin{center}
{\it Theoretical Physics, Oxford University,
1 Keble Rd., Oxford OX1 3NP, UK}
\end{center}
\vskip 1 cm
\centerline{\bf ABSTRACT }
\medskip
The investigation of symmetry nonrestoration scenarios has led
to a controversy, with certain nonperturbative approximation
schemes giving indications in sharp disagreement with those
found within conventional perturbation theory.
A Rayleigh-Ritz variational approach to the problem,
which might be useful in bridging the gap between perturbative
and nonperturbative viewpoints, is here proposed.
As a first application, this approach is used in the investigation
of a $Z_2 \! \times \! Z_2$-invariant thermal field theory
with two scalar fields, placing particular emphasis on
the region of parameter space that has been claimed to support
symmetry nonrestoration.

\vfill
\noindent{OUTP-96-44P \space\space\space 
hep-ph/9610262
\hfill July 1996}

\newpage
\pagenumbering{arabic}
\setcounter{page}{1}
\pagestyle{plain}
\baselineskip 12pt plus 0.2pt minus 0.2pt

The subject of
temperature-induced phase
transitions[1-3]
%%%transitions\cite{linde,doja,wei}
in relativistic quantum field theories
has been extensively investigated
over the last twenty years.
In particular,
transitions
from a high-temperature
symmetric phase to a low-temperature phase in which some symmetries
are spontaneously broken
are a crucial ingredient in most modern cosmological scenarios,
and have been shown to be realized in large classes of
thermal field theories.

The possibility of
symmetry nonrestoration (SNR) at high 
temperatures[3-7]
%%%temperatures\cite{wei,mohase,ictp1,wilsnr,dvalmore}
(or transitions
from a high-temperature
broken-symmetry phase to a low-temperature
symmetric phase\cite{pilang})
could also have interesting phenomenological implications,
most notably allowing to circumvent
the {\it monopole problem}
in certain {\it Grand Unification Theories};
however,
the investigation of SNR scenarios has led
to controversy, with certain nonperturbative approximation
schemes giving indications in sharp disagreement with those
found within conventional perturbation theory.
Specifically, whereas 
perturbative analyses find that some models,
possibly of phenomenological relevance,
can support SNR upon appropriate (however {\it ad hoc})
choice of the available parameters,
the corresponding analyses within
certain nonperturbative approximation schemes\cite{nonpertsnr}
indicate that symmetry is inevitably restored.

Recently, some progress has been made toward
bridging the gap between perturbative and nonperturbative
results on SNR. This has been attained via the use of improved
perturbative techniques\cite{bilo,snr1}
in which, while preserving the general structure
of the perturbative expansion, some nonperturbative 
features of the theory are effectively taken into account.
The preliminary results of this improved
perturbative approaches have indicated\cite{bilo,snr1}
that the conventional (unimproved) perturbative techniques 
overestimate the ``{\it SNR parameter space}''
(the region of parameter space capable
of supporting SNR), 
and it is reasonable to interpret
these results as suggesting that further improvements
in the accuracy of the approximations 
would ultimately lead to the conclusion that 
the {\it SNR parameter space} is actually empty,
just as predicted within the nonperturbative 
approximation schemes adopted in Ref.\cite{nonpertsnr}.
This expectation is encouraged by the findings of related
studies on the lattice\cite{bilo3}.

In this Letter, 
I discuss techniques which can be useful in making further
progress in the direction proposed in the
Refs.\cite{bilo,snr1},
and, as a first application, which also serves as illustrative example,
I use them in the investigation
of the two-scalar-field theory of Euclidean Lagrange density 
\begin{equation}
L = {1 \over 2} (\partial_{\mu} \Phi) (\partial^{\mu} \Phi) +
{1 \over 2} (\partial_{\mu} \Psi) (\partial^{\mu} \Psi) 
+{1 \over 2} m^2 \Phi^2
+{1 \over 2} \omega^2 \Psi^2
+{\lambda_\Phi \over 24}  \Phi^4
+{\lambda_\Psi \over 24}  \Psi^4
-{\lambda_{\Phi \Psi} \over 4}  \Phi^2 \Psi^2
~,
\label{lfzz}
\end{equation}
which is $Z_2 \! \times \! Z_2$ 
invariant [$(\Phi \rightarrow -\Phi) \times (\Psi \rightarrow -\Psi)$],
and is among the strongest candidates as a model
supporting SNR.

Within conventional perturbation theory
the model (\ref{lfzz}) is found to support SNR
when\footnote{Note that here and in the following
$\lambda_\Phi$ and $\lambda_\Psi$ are assumed to be positive,
so that there is at least a chance of having a stable theory,
and all the $\lambda$'s are
assumed to be small, so that perturbation 
theory has at least a chance
of giving consistent results.}
\begin{equation}
\lambda_\Phi < \lambda_{\Phi \Psi} < \sqrt{\lambda_\Phi \lambda_\Psi} 
\label{cond}
\end{equation}
or
\begin{equation}
\lambda_\Psi < \lambda_{\Phi \Psi} < \sqrt{\lambda_\Phi \lambda_\Psi} 
~.
\label{condbis}
\end{equation}
The lower bound on $\lambda_{\Phi \Psi}$ in the above (\ref{cond})
and (\ref{condbis}) is set by the requirement that,
in the high temperature limit,
one of the fields
have imaginary
dressed (thermal) mass, which 
within perturbation theory is found to be given
by $M \sim \sqrt{(\lambda_{\Phi} - \lambda_{\Phi \Psi}) T^2}$
and
$\Omega \sim \sqrt{(\lambda_{\Psi} - \lambda_{\Phi \Psi}) T^2}$
respectively for $\Phi$ and $\Psi$.
(Symmetry is not restored if 
the square 
of the dressed mass, {\it i.e.} 
the second derivative
of the effective potential at the origin, is negative.)
The common upper bound on $\lambda_{\Phi \Psi}$
given in (\ref{cond})
and (\ref{condbis}) follows from (naive\cite{snr1}) 
stability analysis\cite{wei}.
However, in light
of the improved perturbative analyses of Refs.\cite{bilo,snr1}
it appears that
conventional perturbation theory
underestimates the lower bound of the {\it SNR parameter space},
since the inclusion of higher order effects 
shows\cite{bilo,snr1}
that the condition
$\lambda_\Phi \! < \! \lambda_{\Phi \Psi}$ 
(or equivalently $\lambda_\Psi \! < \! \lambda_{\Phi \Psi}$)
is not sufficient to render one of the dressed masses imaginary,
and might also
overestimate the upper bound of the {\it SNR parameter space},
since the higher-order
analysis of ultraviolet structures 
indicates\cite{snr1} that the 
condition $\lambda_{\Phi \Psi} < \sqrt{\lambda_\Phi \lambda_\Psi}$
is necessary but not sufficient for stability.

While clearly a lot could be learned by 
further extending/improving the analyses 
presented in Refs.\cite{bilo,snr1},
this objective is presented with serious 
technical hurdles, primarily as a result of the necessity
to study
complicated self-consistent
equations for the dressed masses (which pick up a
momentum dependence beyond leading order).
In this Letter, I use a variational approximation
to estimate the leading correction to the 
results of Refs.\cite{bilo,snr1}.
Specifically, I am interested in the way this
leading correction modifies the result for the lower bound
of the {\it SNR parameter space}. If a more stringent
bound was found, one would be presented with a scenario
in which each next step of improvement of the analysis
appears to lead to a smaller {\it SNR parameter space},
pointing toward the validity of the above mentioned conjecture
that the {\it SNR parameter space} is actually empty.
However, as reported below,
I find that instead the leading correction to the 
result of Refs.\cite{bilo,snr1} 
on the lower bound
of the {\it SNR parameter space}
is {\it in favour}
of SNR.
This suggests that 
additional steps of improvement of the analysis
might lead to alternating contributions,
ultimately resulting in a lower bound
of the {\it SNR parameter space}
which does not differ much 
from the one obtained in Refs.\cite{bilo,snr1},
and therefore is still consistent with SNR
in the $Z_2 \! \times \! Z_2$ model under consideration.

Since the analysis here presented is purely concerned with the 
lower bound of the {\it SNR parameter space},
my result does not affect the possibility that the {\it SNR parameter
space} might be found to be empty
once the issues of stability (which are relevant for
the upper bound of the {\it SNR parameter space})
raised in Ref.\cite{snr1} are fully investigated.
This however appears to require several nontrivial 
steps, starting with the identification
(within a study
of the type reported in Ref.\cite{mosh} for the $Z_2$ model)
of the conditions rigorously necessary (and sufficient) for
the stability of the $Z_2 \! \times \! Z_2$ model.
This challenging program is left for future work.

In preparation
for the derivation of the announced result on
the lower bound
of the {\it SNR parameter space},
let me start by revisiting the results
of Refs.\cite{bilo,snr1}. 
I do this in the framework of
the CJT formalism of the effective potential for composite
%%%operators\cite{snr1,corn,pap7,jackbanf},
operators[11,14-16],
which is ideally suited for a variational
analysis of the type I discuss here.
I choose to work in the 
imaginary time formalism of finite temperature
field theory, wherein the fourth component of
momentum is discretized, $k_4 = i \pi n T$
($n$ even for bosons), and I adopt the notation
\begin{equation} 
\hbox{$\sum$}\!\!\!\!\!\!\!\int_k ~ \equiv \,
T \sum^{\infty}_{n=-\infty} \int {d^3k \over (2 \pi)^3}
~.
\label{imfeynb}
\end{equation}
I also exploit the fact that the two
SNR scenarios
emerging within
conventional perturbation theory,
(\ref{cond}) and (\ref{condbis}),
are simply related by renaming of fields and couplings, and
therefore, for the purpose of the analysis here presented,
it is sufficient to consider $\lambda$'s such that
\begin{equation}
\lambda_\Phi < \lambda_{\Phi \Psi} << \lambda_\Psi
~,
\label{case}
\end{equation}
which allows to investigate
the lower bound
of the {\it SNR parameter space}
in the candidate
SNR scenario (\ref{cond}).
This is useful since (\ref{case}) implies\cite{snr1} that
all the interesting structures\footnote{Clearly,
the setup (\ref{case}) implies that at high temperatures
the minimum of the effective potential along the $\psi$
direction is at $\psi \! = \! 0$.}
of the effective potential $V(\phi,\psi)$, corresponding to the
shifts $\{ \Phi,\Psi \} \rightarrow \{ \Phi+\phi,\Psi+\psi \}$,
are to be found in its
projection on the $\psi \! = \! 0$ axis.
Therefore, in the following I concentrate 
on $V(\phi,\psi \! = \! 0)$, {\it i.e.} 
shifts $\{ \Phi,\Psi \} \rightarrow \{ \Phi+\phi,\Psi \}$,
and, for short, I adopt the notation $V(\phi)$
for $V(\phi,\psi \! = \! 0)$.

In Ref.\cite{snr1} this potential $V(\phi)$ was investigated 
within the ``bubble approximation'', which 
is the lowest nontrivial 
order[14-18]
%%%order\cite{corn,pap7,jackbanf,consolicjt,pap8}
of approximation of the effective
potential in the CJT formalism. 
The result of Ref.\cite{snr1}
can be written, using
the (renormalized) gap equations,
as\footnote{Since, for the reasons
discussed above, I am considering only 
the dependence on higher order terms
of the lower bound 
of the {\it SNR parameter space}, which is quite
decoupled\cite{snr1} from the issues of stability and ultraviolet
structure,
for brevity my analysis here does not provide 
details on renormalization (which the
reader can find in Ref.\cite{snr1}),
and also ignores the small cut-off dependent 
contribution to the effective potential
that is present when the theory, as appropriate for a {\it trivial}
theory\cite{snr1,pap7}, is considered with a finite cut-off\cite{snr1}.}
\begin{eqnarray}
V \! \! &=& \! \! 
{{m}^2 \over 2} \phi^2 +
{{\lambda}_\Phi \over 24} \phi^4
+ Q[M]
+ Q[\Omega]
+ ({{m}^2 \over 2} +
{{\lambda}_\Phi \over 4} \phi^2 - {M^2 \over 2})
P[M]
\nonumber\\
& & 
+ ({{\omega}^2 \over 2} +
{{\lambda}_{\Phi \Psi} \over 4} \phi^2 - {\Omega^2 \over 2})
P[\Omega]
- {{\lambda}_\Phi \over 8} 
(P[M])^2
- {{\lambda}_\Psi \over 8} 
(P[\Omega])^2
+ {{\lambda}_{\Phi \Psi} \over 4} 
P[M]
P[\Omega]
~, \label{vrensumaiii}
\end{eqnarray}
where $M$ and $\Omega$ are
the dressed masses of the fields $\Phi$ and $\Psi$
respectively, and are determined by the gap equations
\begin{eqnarray}
M^2 &=& m^2 + {\lambda_\Phi \over 2} \phi^2 
+ {\lambda_\Phi \over 2} P[M] 
- {\lambda_{\Phi \Psi} \over 2} P[\Omega]
~,
\label{gapta}\\
\Omega^2 &=& \omega^2 
+ {\lambda_{\Phi \Psi} \over 2} \phi^2 
+ {\lambda_\Psi \over 2} P[\Omega]
- {\lambda_{\Phi \Psi} \over 2} P[M] 
~.
\label{gaptb}
\end{eqnarray}
In the above (\ref{vrensumaiii})-(\ref{gaptb}),
$P$ is the ``tadpole"\cite{doja}
\begin{eqnarray}
P[X] \equiv 
\hbox{$\sum$}\!\!\!\!\!\!\!\int_p ~ \! 
{1 \over k^2 \! + \! X^2} 
= {X^2 \over 16 \pi^2} 
\ln {X^2 \over \mu^2} 
- \int \! {d^3k \over (2 \pi)^3}~ \! 
\left[\sqrt{|{\bf k}|^2 \! + \! X^2}  
\left( \!  1 \!  
- \!  exp \left( { \sqrt{|{\bf k}|^2 \! 
+ \! X^2} \over T} \right) \! \right) \! \right]^{-1} ,
\label{gxxb}
\end{eqnarray}
and $Q$ is the ``one loop''\cite{doja}
\begin{eqnarray}
Q[X] = \hbox{$\sum$}\!\!\!\!\!\!\!\int_k \, 
\ln [k^2+X^2] \!\! =
\!\! {X^4 \over 64 \pi^2}             
[\ln {X^2 \over \mu^2} - {1 \over 2}] 
+ T \int {d^3k \over (2 \pi)^3}~
\ln \left[ 1- 
exp \left( { \sqrt{|{\bf k}|^2 + X^2} \over T} \right) \right]
~.
\label{voneTreg}
\end{eqnarray}

By stopping at the level of the bubble approximation used
in Refs.\cite{snr1} (and effectively used
in Refs.\cite{bilo} although 
the formalism adopted there is different)
one ends up ignoring contributions 
to the second 
derivative with respect to $\phi$
of the effective potential at the origin ({\it i.e.}
the square 
of the dressed mass
of the field $\Phi$ at $\phi \! = \! 0$)
which are
of order $\lambda^2 \ln \lambda$.
Taking this into account, and using
the following known\cite{doja} ``high temperature''
(small $X/T$) expansions
\begin{eqnarray}
P[X] \!\! &\simeq& \!\! 
{T^2 \over 12} - {X T \over 4 \pi}
- {X^2 \over 16 \pi^2} \ln {X^2 \over T^2} 
~,
\label{pexp}
\end{eqnarray}
\begin{eqnarray}
Q[X] \!\! &\simeq& \!\! - {\pi^2 T^4 \over 90} 
+ {X^2 T^2 \over 24} - {X^3 T \over 12 \pi}
- {X^4 \over 64 \pi^2} \ln {X^2 \over T^2} 
~,
\label{pqexp}
\end{eqnarray}
one finds that the Eqs.(\ref{vrensumaiii})-(\ref{gaptb})
lead to the following result for the
second derivative
of the effective potential at the origin
\begin{eqnarray}
V''(0) &\simeq& {m}^2 
+ (\lambda_\Phi - \lambda_{\Phi \Psi})
{T^2 \over 24}
+ \lambda_{\Phi \Psi}
\sqrt{\lambda_\Psi - \lambda_{\Phi \Psi}}
{T^2 \over 16 \sqrt{6} \pi}
~,
\label{gaphight}
\end{eqnarray}
where indeed 
terms of order $\lambda^2 \ln \lambda$ and higher
have been dropped, and the relations
(\ref{case}) have been used in extracting the leading
contribution of order $\lambda \sqrt{\lambda}$.
Notice that, in the scenario (\ref{case}), the 
contribution to (\ref{gaphight})
of order $\lambda$, which can be estimated in
conventional (unimproved) perturbation theory,
is negative, {\it i.e.} works in favour of SNR,
but the contribution of order $\lambda \sqrt{\lambda}$,
whose evaluation requires to take into account
the bubble diagrams\cite{snr1}, is positive.
Most importantly the positivity
of the correction
of order $\lambda \sqrt{\lambda}$
renders insufficient for SNR
the condition
$\lambda_{\Phi \Psi} \! > \! \lambda_\Phi$,
which sets the lower bound 
of the {\it SNR parameter space}
within conventional perturbation theory.
This is the basis for the recent observation\cite{bilo,snr1}
that the bubble-improved perturbation theory
predicts a smaller {\it SNR parameter space}
than conventional perturbation theory.

The improvement in the determination
of the {\it SNR parameter space} that I report in this Letter
relies on an estimate of the leading
correction
to (\ref{gaphight}),
which is of order $\lambda^2 \ln \lambda$
(notice 
that $|\epsilon|^2 \! < \! |\epsilon|^2 \ln |\epsilon| \! < \! |\epsilon|$
for small $\epsilon$).
I obtain this result via the use of variational
techniques in the framework
of the CJT formalism, which allows to go 
systematically[11,14-16]
%%%systematically\cite{snr1,corn,pap7,jackbanf}
beyond the bubble approximation.
In the CJT formalism the 
effective potential $V$ is obtained
as the solution of a variational problem for the
effective potential for composite operators $W$:
\begin{equation}
V(\phi) = W[\phi;M(\phi;k)]
~,\label{vw}
\end{equation}
\begin{equation}
\biggl[{\delta W[\phi;{\cal M}(k)] \over \delta 
{\cal M}(k)}\biggr]_{{\cal M}(k)=M(\phi;k)}
= 0
~,\label{hfc}
\end{equation}
A rigorous definition of $W$ can be found 
%%%in Refs.\cite{corn,pap7,jackbanf,pap8},
in Refs.[14-17], 
but for the purposes of the
present Letter it is sufficient to observe that $W$ admits a
loop expansion, with ${\cal M}(k)$ 
appearing as the (dressed) mass matrix\footnote{In the present 
Letter the formalism is set up in
terms of the dressed masses, whereas in Ref.\cite{snr1}
it was set up in terms of the dressed propagators. 
The two setups are equivalent
(if no assumption is made
concerning the momentum dependence of the dressed objects),
but the discussion in terms of the dressed masses
guides more directly toward the type of Rayleigh-Ritz approximation
to be discussed later.}:
\begin{eqnarray}
W =
V_{tree}(\phi)
+{1 \over 2} \, 
\hbox{$\sum$}\!\!\!\!\!\!\!\int_k \, 
Tr \ln [k^2+{\cal M}^2] 
+ {1 \over 2} \,  
\hbox{$\sum$}\!\!\!\!\!\!\!\int_k \, 
{m_{tree}^2 - {\cal M}^2 \over
k^2 + [{\cal M}^2(k)] }
+  W^*[\phi;{\cal M}(k)]
~,\label{vsumazzt}
\end{eqnarray}
where $V_{tree}(\phi)$ is the tree-level (classical) potential,
$m_{tree}$ is the tree-level mass, and
$W^*$ is given by all the two-particle-irreducible\cite{corn}
vacuum-to-vacuum graphs with two or more loops
in the theory with vertices given by the interaction part of the 
shifted ($\Phi \rightarrow \Phi + \phi$) Lagrangian and
propagator set equal to $G(k)$,
with $[G^{-1}(k)]_{ab} = \delta_{ab}  k^2 + [{\cal M}^2(k)]_{ab}$.

The bubble approximation, which 
is the lowest nontrivial 
order[14-18]
%%%order\cite{corn,pap7,jackbanf,consolicjt,pap8}
of approximation of the effective
potential in the CJT formalism,
is obtained 
by including in $W^*$ 
only the ``double-bubble diagrams'', {\it i.e.}
diagrams made of two rings touching at one point.
Importantly, the contributions of double-bubble diagrams have the form
\begin{eqnarray}
V_{\bigcirc \! \bigcirc} (X,Y) &=& 
\hbox{$\sum$}\!\!\!\!\!\!\!\int_k \, 
\hbox{$\sum$}\!\!\!\!\!\!\!\int_p ~
{1 \over k^2 + X^2} 
{1 \over p^2 + Y^2} = P[X] P[Y]
~,\label{bubblestructure}
\end{eqnarray}
which does not involve any flow of momenta from one loop to the other.
The possibility to make substantial analytic progress,
ultimately leading to the result (\ref{vrensumaiii}),
is a peculiarity of the bubble approximation,
which in particular 
involves dressed masses that are exactly momentum
independent\cite{snr1,pap8}.
As soon as one goes beyond the bubble approximation
the analysis becomes much more complicated.
The dominant higher-order contributions,
already relevant at the
order $\lambda^2 \ln \lambda$
here under consideration,
come from ``sunset'' diagrams
\begin{eqnarray}
V_{\bigcirc \!\!\!\!\!\! - \!\!\!\,\! -}(X,Y,Z) =
\hbox{$\sum$}\!\!\!\!\!\!\!\int_k \, 
\hbox{$\sum$}\!\!\!\!\!\!\!\int_p ~
{1 \over k^2 + X^2} 
{1 \over p^2 + Y^2} 
{1 \over (k+p)^2 + Z^2} 
~,\label{sunsetstructure}
\end{eqnarray}
which arise from the contraction of two copies of any
three-point vertex present in the shifted
($\{ \Phi,\Psi \} \rightarrow \{ \Phi+\phi,\Psi \}$)
Lagrangian.
In these diagrams momentum does flow from one loop to the other
leading to the loss of all simplifications
encountered in the bubble approximation.
In such cases, analytic\footnote{The variational problem
in which the CJT formalism casts the evaluation of the
effective potential is well suited\cite{pap8} for exact numerical
analysis even when non-bubble contributions are included,
although, as a consequence of triviality,  some care is needed
in the handling of ultraviolet
structures.}
progress requires that the effective potential
be evaluated approximately.
The fact that in the CJT formalism the
effective potential is obtained as the solution
of a variational problem renders available the ``machinery''
of variational approximations.
In particular, one can resort to approximations\footnote{The reader
should keep in mind that there are two levels of approximation
involved in the analysis presented in this Letter.
To begin with, I am working within truncations/approximations
of the effective potential that arise within the
formalism of the loop expansion of the CJT effective potential.
When this loop expansion is truncated at the ``bubble level''
the calculation can be completed without further approximation.
If one goes beyond the bubble approximation, to the ``sunset level'',
even the truncated expression of the CJT effective potential
cannot be analyzed exactly (unless numerical techniques are 
developed), and a further variational approximation
is needed to estimate analytically the
sunset-truncated CJT effective potential.}
of the 
Rayleigh-Ritz type\cite{corn,pap8}, in which,
rather than considering arbitrary variations of the dressed
masses ${\cal M}_{ab}(k)$,  
one takes a parameter-dependent {\it ansatz} 
for ${\cal M}_{ab}(k)$ and vary only the parameters.
This variational approach was recently used in Ref.\cite{pap8}
in a related study of gauge theories at finite temperature,
and has been well received\cite{zwirn,shapo}, although 
it is hard\cite{pap8,zwirn} 
to establish its range of validity
in the framework of the (``thermally troublesome'') gauge theories.
The non-gauge-theoretical model
considered in the present Letter
is, however, closer to the contexts
in which Rayleigh-Ritz variational approaches
have been traditionally used, and one can expect
the usual arguments\footnote{Discussion of these topics can be found
in Ref.\cite{betjack}.}
for its reliability to hold.
Therefore, 
a Rayleigh-Ritz variational approach can be
useful in investigations of the issue
of restoration/nonrestoration
of symmetry in the $Z_2 \! \times \! Z_2$ model,
and, as a first step in that direction,
I use it here in estimating the leading
non-bubble contribution
to the effective potential, which comes from the sunset
diagrams.

Before proceeding with this estimate, let me observe
that within the bubble approximation
one could have already resorted to a Rayleigh-Ritz approximation.
In fact, 
as the reader can easily check,
by substituting 
in the bubble-approximated CJT effective potential 
for composite operators
the following variational {\it ansatz}
for the dressed masses
\begin{eqnarray}
{\cal M}_{ab} &=& 
\delta_{a1} \delta_{b1} M +
\delta_{a2} \delta_{b2} \Omega
~,\label{bubblerr}
\end{eqnarray}
one obtains a Rayleigh-Ritz bubble effective potential
which, after stationarizing with respect to the (dimensionful) 
parameters $M$ and $\Omega$,
reproduces Eq.(\ref{vrensumaiii})
within the accuracy achievable in the bubble approximation,
with $M$ and $\Omega$
identified with the 
solutions of (\ref{gapta})-(\ref{gaptb}).
In fact this is encoded in the fact that
the bubble effective potential obtained in Ref.\cite{snr1}
is stationary 
(again up to terms which are however
not reliably determined by the bubble approximation)
with respect to variations of $M$ and $\Omega$
in the neighborhood of the solutions
of the gap equations (\ref{gapta})-(\ref{gaptb}).
In this sense the Rayleigh-Ritz variational approach is exact
within the bubble approximation, as it should be expected
since the variational {\it ansatz} (\ref{bubblerr})
is just in the form of the exact (bubble-approximated) dressed mass
matrix.

In order to estimate the contribution of the sunset
diagrams within the Rayleigh-Ritz approach 
it is sufficient to take the following three steps.
First, add the relevant sunset terms to 
the bubble-approximated CJT effective potential. 
Then, evaluate the resulting expression for the
effective potential with a physical parameter-dependent {\it ansatz}
for the dressed masses.
Finally, stationarize with respect to the parameters characterizing 
the {\it ansatz}.

Clearly in this higher-order analysis the result of
the Rayleigh-Ritz approach is not exact,
but, through the variational procedure, it should provide
a reliable\cite{betjack} way to 
encode into a handleable set of parameters
the bulk of the effect of the sunset-induced
momentum dependence of the dressed masses.
Actually, as a result of the observation above
concerning the exactness of the {\it ansatz} (\ref{bubblerr})
at the bubble level,
one can expect that the {\it ansatz} (\ref{bubblerr})
be sufficient also to estimate the leading sunset correction;
in fact, as long as
the bubble contributions lead over the
sunset ones
({\it i.e.} at small values of the coupling constants)
the dressed masses will be only weakly momentum dependent,
so that it should be possible to replace them with
momentum independent ``effective masses" to be determined
self-consistently.

Adopting the {\it ansatz} (\ref{bubblerr})
one finds that the handling of sunset diagrams
only requires to
evaluate $V_{\bigcirc \!\!\!\!\!\! - \!\!\!\,\! -}(X,Y,Z)$
with momentum independent $X$, $Y$, and $Z$.
This has already been done in Refs.\cite{parwani,arnold},
where, in particular, it was shown that the leading
high temperature contribution, which is
of interest for my analysis,
is given by
\begin{eqnarray}
V_{\bigcirc \!\!\!\!\!\! - \!\!\!\,\! -}(X,Y,Z) \simeq {T^2 \over 
16 \pi^2} \ln \left( {3 T \over X+Y+Z} \right)
~.
\label{infraredteta}
\end{eqnarray}

From Eq.(\ref{vsumazzt}) one can easily show
that the complete
``sunset-approximated" ({\it i.e.} obtained by adding
the contributions from the
sunset diagrams to the
bubble-approximated effective potential)
CJT effective potential can be written as 
\begin{eqnarray}
V \! \! &=& \! \! 
{{m}^2 \over 2} \phi^2 +
{{\lambda}_\Phi \over 24} \phi^4
+ Q[M]
+ Q[\Omega]
+ ({{m}^2 \over 2} +
{{\lambda}_\Phi \over 4} \phi^2 - {M^2 \over 2})
P[M]
\nonumber\\
& & 
+ ({{\omega}^2 \over 2} +
{{\lambda}_{\Phi \Psi} \over 4} \phi^2 - {\Omega^2 \over 2})
P[\Omega] + {{\lambda}_\Phi \over 8} 
(P[M])^2
+ {{\lambda}_\Psi \over 8} 
(P[\Omega])^2
- {{\lambda}_{\Phi \Psi} \over 4} 
P[M]
P[\Omega] \nonumber\\
& & - {\lambda_\Phi^2 \over 12} \phi^2
V_{\bigcirc \!\!\!\!\!\! - \!\!\!\,\! -}(M,M,M)
- {\lambda_{\Phi \Psi}^2 \over 4} \phi^2
V_{\bigcirc \!\!\!\!\!\! - \!\!\!\,\! -}(\Omega,\Omega,M)
~. \label{rrvcjt}
\end{eqnarray}
As discussed above, the ordinary
effective potential
is obtained from this formula by
stationarizing with respect to variations of $M$ and $\Omega$.
Then, also using the high temperature expansions (\ref{pexp}),
(\ref{pqexp}), and
(\ref{infraredteta}), 
with some simple algebra one can show that
the sought contribution\footnote{Terms
of order $\lambda^2 \ln \lambda$
are already present in the bubble approximation
(these terms are among those listed in Eq.(\ref{logs})),
but could not be included\cite{bilo}
in the bubble approximation of the 
second derivative of the effective potential at the origin,
given in Eq.(\ref{gaphight}),
since they are not reliably determined when the sunset contributions
are not taken into account.}
of order $\lambda^2 \ln \lambda$ to
the second derivative of the effective potential at the origin
is given by
\begin{eqnarray}
&&{ \lambda_{\Phi \Psi} \Omega^2 \over
16 \pi^2} \ln ({\Omega \over T})
- {\lambda_{\Phi} M^2 \over 16 \pi^2} \ln ({M \over T})
+{ \lambda_{\Phi \Psi}^2 T^2 \over 16 \pi^2} \ln ({2 \Omega
+ M \over 3 T})
+ { \lambda_{\Phi}^2 T^2 \over 48 \pi^2} \ln ({\Omega \over T}) \simeq
\nonumber\\
&& ~~~~~~~~~~~~~~~~~~~~~~~~~~~~~
\simeq { \lambda_{\Phi \Psi} \Omega^2 \over
16 \pi^2} \ln ({\Omega \over T})
\simeq - { \lambda_{\Phi \Psi} \lambda_{\Psi} T^2 \over 768 \pi^2}
\ln {24 \over \lambda_\Psi} < 0 
~, 
\label{logs}
\end{eqnarray}
where
the indicated approximations hold
in the scenario (\ref{case}) presently under consideration,
which also implies $M < \! < \Omega$.

Eq.(\ref{logs}) shows that,
as anticipated in the opening of this Letter,
the leading $O(\lambda^2 \ln \lambda)$
correction to Eq.(\ref{gaphight})
is negative, {\it i.e.} works in favour of SNR.
This is certainly of encouragement for the possibility
of SNR, with the proviso about stability mentioned
in the opening of this Letter.
It should also be noticed that the SNR-favouring correction
here identified is quite small, especially as a result
of the numerical prefactor, and, while possibly sufficient
to help the case of SNR in the toy model here considered,
it might be negligible in the context of
models relevant for particle physics, where large
gauge-coupling contributions render SNR very unlikely\cite{bilo}.
The Rayleigh-Ritz variational approach here advocated
should be useful in further investigation of these issues.
 
\bigskip
\bigskip
\bigskip
It is a pleasure to acknowledge conversations with 
G. Bimonte, R. Jackiw, G. Lozano,
O. Philipsen, S.-Y. Pi, and S. Sarkar.

\newpage
\baselineskip 12pt plus .5pt minus .5pt


\begin{thebibliography}{99}
\bibitem{linde} D. Kirzhnits, JETP Lett. 15 (1972) 529;
D. Kirzhnits and A. Linde, Phys. Lett. B42 (1972) 471;
D. Kirzhnits and A. Linde, JETP 40 (1974) 628.
\bibitem{doja} L. Dolan and R. Jackiw, Phys. Rev. D 9 (1974) 3320.
\bibitem{wei} S. Weinberg, Phys. Rev. D9 (1974) 3357.
\bibitem{mohase} R.N. Mohapatra and G. Senjanovic,
Phys. Rev. Lett. 42 (1979) 1651;
Phys. Rev. D20 (1979) 3390. 
\bibitem{ictp1} G. Dvali, A. Melfo, and
G. Senjanovic, Phys. Rev. Lett. 75 (1995) 4559.
\bibitem{wilsnr} T. G. Roos, 
hep-th/9511073.
\bibitem{dvalmore} G. Dvali, A. Melfo, and G. Senjanovic,
hep-ph/9601376;
G. Dvali and K.Tamvakis,
hep-ph/9602336.
\bibitem{pilang} P. Langacker and S.-Y. Pi,
Phys. Rev. Lett. 45 (1980) 1. 
\bibitem{nonpertsnr} Y. Fujimoto and S. Sakakibara, 
Phys. Lett.  B151 (1985) 260;
E. Manesis and S. Sakakibara, Phys. Lett. B157 (1985) 287;
G.A. Hajj and P.N. Stevenson, Phys. Rev. D413 (1988) 413;
K.G.Klimenko, Z.Phys. C43 (1989) 581; Theor. Math. Phys. 80 (1989) 929.
\bibitem{bilo} G. Bimonte and G. Lozano, Phys. Lett. B366 (1996) 248;
Nucl. Phys. B460 (1996) 155.
\bibitem{snr1} G. Amelino-Camelia, Nuc. Phys. B476 (1996) 255.
\bibitem{bilo3} G. Bimonte and G. Lozano, hep-th/9603201.
\bibitem{mosh} W.A. Bardeen and Moshe Moshe,
Phys. Rev. D28 (1983) 1372.
\bibitem{corn} J.M. Cornwall, R. Jackiw, and E. Tomboulis,
Phys. Rev. D10, 2428 (1974).
\bibitem{pap7} G. Amelino-Camelia and S.-Y. Pi, Phys. Rev. D47 (1993) 2356.
\bibitem{jackbanf} R. Jackiw and G. Amelino-Camelia, hep-ph/9311324,
in {\em Proceedings of the Third Workshop on Thermal Field Theories
and Their Applications, Banff, Canada, August 15-27, 1993},
edited by F.C. Khanna, R. Kobes, G. Kunstatter, and
H. Umezawa (World Scientific, 1994).
\bibitem{consolicjt}P. Castorina, M. Consoli, and D. Zappala,
Phys. Lett. B201 (1988) 90. 
\bibitem{pap8} G. Amelino-Camelia, Phys. Rev. D49 (1994) 2740.
\bibitem{zwirn} J.R. Espinosa, M. Quiros, 
and F. Zwirner, Phys. Lett. B314 (1993) 206. 
\bibitem{shapo} K. Farakos, K. Kajantie, K. Rummukainen, 
and M. Shaposhnikov, Nucl. Phys. B425 (1994) 67.
\bibitem{betjack} H.A. Bethe and R. Jackiw,
{\it Intermediate Quantum Mechanics}, third ed. (The Benjamin/Cummings
Publishing Company,
Menlo Park, California, 1986).
\bibitem{parwani} R.R. Parwani, 
Phys. Rev. D45 (1992) 4695.
\bibitem{arnold} P. Arnold and O. Espinosa, 
Phys. Rev. D47 (1993) 3546.

\end{thebibliography}
\end{document}